\documentclass[showpacs]{revtex4}
\usepackage{graphicx,latexsym}

\begin{document}
\draft
\title{All-optical versus electro-optical quantum-limited feedback}
\author{H.M. Wiseman and G.J. Milburn}
\address{Department of Physics, University of Queensland,
Queensland 4072 Australia}
\date{Phys. Rev. A {\bf 49}, 4110 (1994).}

\begin{abstract}
All-optical feedback can be effected by putting the output of a source
cavity through a Faraday isolator and into a second cavity which is
coupled to the source cavity by a nonlinear crystal. If the driven
cavity is heavily damped, then it can be adiabatically eliminated and a
master equation or quantum Langevin equation derived for the first
cavity alone. This is done for an input bath in an arbitrary state,
and for an arbitrary nonlinear coupling. If the intercavity coupling
involves only the intensity  (or one quadrature) of the driven cavity,
then the effect on the source cavity is identical to that which can be
obtained from electro-optical feedback using direct (or homodyne)
detection. If the coupling involves both quadratures, this equivalence
no longer holds, and a coupling linear in the source amplitude can 
produce a nonclassical state in the source cavity. The analogous 
electro-optic scheme using heterodyne detection introduces extra 
noise which prevents the production of nonclassical light. Unlike the 
electro-optic case, the all-optical feedback loop has an output beam 
(reflected from the second cavity). We show that this may be 
squeezed, even if the source cavity remains in a classical state.
\end{abstract}


\maketitle

\section{Introduction}

Electro-optic feedback is the use of a photocurrent to control the 
source of the light incident on the detector producing that current.
Such feedback has long been used to control noise in optical systems 
such as lasers. In many cases, the noise which is being controlled is 
classical noise. That is, noise which is well above the shot-noise limit 
(also known as the quantum limit). However, technological advance 
now enables experimentalists to work with light which is at, or even 
(thanks to squeezing \cite{JOSABSS87}) below, the quantum limit.  
Thus, the quantum limits to electro-optic feedback have been of 
considerable interest in the past decade 
\cite{YamImoMac86,HauYam86,Sha87,TapRarSat88}. Our interest 
lies with feedback onto the dynamics of the source cavity 
\cite{WisMil93b,WisMil93d,Wis93b}. We have described such 
feedback using quantum measurement theory, in the form of 
stochastic quantum trajectories 
\cite{Car93b,DalCasMol92,GarParZol92,WisMil93c}. 
In the limit that the time delay 
in the feedback loop is much less that the cavity lifetime, a master 
equation describing the feedback can be derived. This is a simple and 
elegant way to treat feedback. There is also a corresponding Langevin 
equation approach \cite{Wis93b}, which is more convenient for some 
purposes. The clearest result of our theory is that, unless the system 
without feedback has nonclassical dynamics, then controlling its 
dynamics via a photocurrent cannot produce nonclassical light. That 
is, feedback based on external photodetection cannot produce 
squeezing.

Given these limitations on the noise-reduction abilities of electro-optic 
feedback, it is natural to ask, what about all-optical feedback. That is, 
instead of detecting the emitted light, it is reflected around a loop back 
into the source cavity, or, more fruitfully, into another cavity which is 
coupled to the first cavity in some way. Of course, for this mechanism 
to be considered feedback, the feedback loop must be one-way, 
otherwise we would simply be describing a pair of doubly-coupled 
cavities. The general configuration under consideration is shown 
schematically in Fig.~1. One mechanism for achieving the required 
unidirectionality is the Faraday isolator which utilizes Faraday 
rotation and polarization-sensitive beam splitters.  A quantum 
theoretical treatment which incorporates this spatial symmetry 
breaking at the level of the Hamiltonian  has been given  recently by 
Gardiner \cite{Gar93}. If the propagation time between the source 
system the and driven  system is ignored, then a master equation for 
both systems may be derived. This result was obtained simultaneously 
by Carmichael \cite{Car93a}, who introduced the term ``cascaded 
systems". The new feature which we consider here is for the driven 
system to be coupled back to the source system via some interaction 
Hamiltonian, giving rise to all-optical feedback.

\begin{figure}
\caption{[See end.] Diagram of the general all-optical feedback scheme 
considered. The annihilation operators for the source and driven 
cavities 
are denoted $c_1$ and $c_2$ respectively, while $b_2$ and $b_3$ 
represent traveling waves. The nonlinear coupling 
between the cavities is indicated by $V$, $FR$ denotes a Faraday 
rotator, and $PBS$ a polarization-sensitive beam splitter. }
\end{figure}

The quantum theory of cascaded systems is summarized in Sec.~II, 
and the simplest example of feedback via this method is considered. 
In Sec.~III, we consider a more complex feedback model utilizing a 
nonlinear crystal to influence the source cavity, with the strength of 
the feedback being proportional to the intensity of the source cavity.  
Both a master equation and a quantum Langevin 
equation are derived, and are shown to be approximately the same as 
those pertaining to electro-optic feedback via direct detection. 
Similarly, the quadrature-sensitive all-optical feedback scheme 
considered in Sec.~IV is equivalent to homodyne-mediated feedback 
in its effect on the source cavity. It is thus not possible to produce a 
nonclassical state in the source cavity via a coupling which is linear in 
the source cavity amplitude or intensity (which correspond to classical 
driving or detuning). 
Unlike the electro-optic case, the all-optical feedback loop has an 
output beam of light (reflected off the driven cavity), and we show 
that this may be squeezed even 
though the source cavity is not. Sec.~V treats an all-optical feedback 
scheme which is sensitive to both quadratures of the source cavity. 
This has no direct electro-optic equivalent, and may produce a 
nonclassical source state even with a coupling linear in the source 
amplitude. An electro-optic analogue can be defined, and the extra 
noise which it introduces  seen explicitly. The two schemes are 
contrasted using the simplest feedback example introduced in Sec.~II. 

\section{Cascaded systems}

\subsection{Unidirectional coupling}
In this section, we present a summary of the theory of cascaded open 
quantum systems. For different presentations, see 
Refs.~\cite{Gar93,Car93a}. Our starting point is the input-output 
theory of open quantum systems \cite{GarCol85,Car87}. This theory 
describes a system interacting locally with a bath consisting of a 
continuum of harmonic oscillators. Physically, the system may be an 
optical cavity, and the bath the external electromagnetic field  modes 
with momentum aligned to the cavity axis. The electric field (or 
rather, one polarization component) at a 
particular point in space-time (paramaterized by $z,t$) is represented 
approximately by the Heisenberg picture operator \cite{Car87}
\begin{equation}
E(z,t) = \sqrt{\frac{\hbar k}{2 \epsilon_0 A}} \left[ b(z,t) + b^\dagger 
(z,t) 
\right].
\end{equation}
Here, $A$ is the cross sectional area of the beam,  and only 
frequencies near the central wave number $k$ are assumed to be of 
interest. The canonical commutation relations for the 
complex amplitudes $b(z,t)$ are
\begin{equation}
[b(z,t),b^\dagger(z',t)] = c \delta(z-z'), \label{ccr}
\end{equation}
where $c$ is the speed of light, and for regions where the field 
propagates freely,
\begin{equation}
b(z,t+\tau) = b(z-c\tau,t). \label{prop}
\end{equation}

Let the external field be coupled to the cavity by a very good mirror at  
$z=0$. The field with $z<0$ then represents an incoming field and 
that with $z>0$ an outgoing one. We assume a linear coupling of the 
form
\begin{equation}
H_1(t) = i \hbar \sqrt{\gamma_1}[b^\dagger(0,t) c_1(t) - c_1^\dagger 
(t) b (0,t) ],
\end{equation}
where $\gamma_1$ is the coupling constant having the dimension of 
inverse time, and $c_1(t)$ is the annihilation operator of the cavity 
tuned to the frequency $ck$. Ignoring other dynamics, the evolution 
of an arbitrary Heisenberg operator $a(t)$ is 
\begin{equation}
\dot{a}(t) = -[b^\dagger(0,t) c_1(t) - c_1^\dagger (t) b(0,t) , a(t) ].
\end{equation}
Now because of the singularity of the canonical commutation 
relations (\ref{ccr}), it is necessary to be careful in dealing with this 
evolution equation. A convenient method is to use quantum Ito 
stochastic differential calculus \cite{GarCol85,Bar86}. Define an 
input 
field, representing the field just before it interacts with the cavity at 
time 
$t$ by
\begin{equation}
b_1 (t) = b(0^-,t).
\end{equation}
This can be thought of as a white noise term, independent of the state 
of the cavity at time $t$. The analog of the Weiner increment in the 
Ito calculus is then
\begin{equation}
dB_1 (t) = b_1 (t) dt,
\end{equation}
which satisfies
\begin{equation}
[dB_1(t),dB_1^\dagger(t)] = dt.
\end{equation}
The evolution of an arbitrary operator is then given by
\begin{equation}
a(t+dt) = U_1 ^\dagger (t,t+dt) a(t) U_1(t,t+dt) \label{a1}
\end{equation}
where 
\begin{equation}
U_1(t,t+dt) = \exp \left\{ \sqrt{\gamma_1} \left[ dB_1^\dagger (t) 
c_1(t) - 
dB_1 (t) c_1(t) \right] \right\}. \label{U1}
\end{equation}

In Eq.~\ref{a1}, the bath operators $dB_1(t)$ and $dB_1^\dagger(t)$ 
are 
independent of 
the system operator $a(t)$, which makes it an easy equation to solve. 
However, the price which must be paid by this simplification of Ito 
calculus is that $U_1(t,t+dt)$ must be expanded to second order in the 
increment $dB_1(t)$. Now if $b_1(t)$ is to be thought of as a bath, it 
should be specifiable simply by its moments. We need only the first 
and second order moments. The first order moment corresponds to a 
coherent amplitude
\begin{equation}
\langle dB_1(t) \rangle = \beta(t) dt,
\end{equation}
while the second order moments indicate white noise:
\begin{eqnarray}
\langle dB_1^\dagger (t) dB_1(t) \rangle &=& Ndt = \langle dB_1 (t) 
dB^\dagger_1(t) \rangle - 1 \\
\langle dB_1(t) dB_1(t) \rangle &=& Mdt = \langle dB^\dagger_1(t)
dB^\dagger(t) \rangle ^* .
\end{eqnarray}
These equations include the cases of thermal noise ($M=0$), and 
perfectly squeezed white noise [$|M|^2 = N(N+1)$]. It is nonclassical 
only if $|M| > N$. Note that even if $N=0$, it is necessary to expand 
$U_1(t,t+dt)$ to second order because then $dB_1(t) dB_1^\dagger (t) 
= 
dt$, which could be regarded as vacuum noise. The full result 
(with time arguments omitted for convenience) is the quantum 
Langevin equation
\begin{eqnarray}
da &= &\frac{\gamma_1}{2} \left[ (N+1) (2c_1^\dagger a c_1 - a 
c_1^\dagger 
c_1 - c_1^\dagger 
c_1 a) + N (2c_1ac_1^\dagger - ac_1 c_1^\dagger - c_1 
c_1^\dagger a) \right. \nonumber \\
&& \left. + M [c_1^\dagger, [ c_1^\dagger,a]] + M^* 
[c_1,[c_1,a]] \right] dt  - \sqrt{\gamma_1} [ dB_1^\dagger  c_1 - 
dB_1 c_1^\dagger , a] .
\end{eqnarray}

The final, stochastic term in this equation is essential to preserve 
canonical commutation relations. However, the stochastic terms can 
be ignored when changing from the Heisenberg to the Schr\"odinger 
picture and deriving the evolution of the density operator for the 
cavity mode alone. This is found from the relation
\begin{equation}
\langle da (t) \rangle = {\rm Tr} [d\rho(t)\, a],
\end{equation}
where  the picture (Schr\"odinger or Heisenberg) is specified by the 
placement of the time argument. The resulting master equation is
\begin{eqnarray}
\dot{\rho}(t) &=& \gamma_1 \left[ (N+1) {\cal D}[c_1]\rho + N 
{\cal D}[c_1^\dagger]\rho  + \frac{M}{2} 
[c_1^\dagger,[c_1^\dagger,\rho]] + 
\frac{M^*}{2} [c_1,[c_1,\rho]] \right] \nonumber \\
&&+ \sqrt{\gamma_1} [ \beta^*(t) c_1 - \beta (t) c_1^\dagger ,\rho], 
\label{sys1me}
\end{eqnarray}
where we have defined a superoperator ${\cal D}$ taking an arbitrary 
operator as its argument by
\begin{equation}
{\cal D}[a] \rho = a\rho a^\dagger - \frac{1}{2} a^\dagger a \rho - 
\frac{1}{2} 
\rho a^\dagger a.
\end{equation}

It might be thought that this master equation is the only product of the 
above theory which is of any significance, since it generates the 
evolution of the cavity mode. However, experimentally, it is often the 
light which leaves the cavity through the output mirror, rather than the 
internal state of the cavity, which is of interest. This is specifically 
so 
for this paper, in which the output of the cavity will be used in the 
feedback loop. Thus, we need an expression for the field leaving the 
cavity. From Eq.~(\ref{prop}), this is evidently given by
\begin{equation}
b_2(t) \equiv b(0^+,t) = U_1(t,t+dt)^\dagger b_1(t) U_1(t,t+dt).
\end{equation}
To lowest order in $dt$, this is
\begin{equation}
b_2(t) = b_1(t) + \sqrt{\gamma_1} c_1(t).
\end{equation}
Just as $b_1(t)$ is independent of, and so commutes with an arbitrary 
system operator $a(t')$ at an earlier time $t' < t$, the output field 
commutes with all system operators at a later time.

Now let this field be the input into another cavity with annihilation 
operator $c_2(t)$. If the damping rate for this cavity is $\gamma_2$, 
then the Hamiltonian coupling is
\begin{equation}
H_2(t) = i \hbar \sqrt{\gamma_2}[b^\dagger(c\tau,t) c_2(t) - 
c_2^\dagger (t)
 b (c\tau,t) ], \label{H2}
\end{equation}
where $c\tau$ is the path length between the two cavities. Proceeding 
as above, the unitary evolution generated by this Hamiltonian is
\begin{equation}
U_2(t,t+dt) = \exp \left\{ \sqrt{\gamma_2} \left[ dB_2^\dagger (t-\tau) 
c_2(t) - dB_1 (t-\tau)  c_2^\dagger(t) \right] \right\}, \label{U2}
\end{equation}
where $dB_2(t) = b_2(t)dt$.  An arbitrary operator in the source or 
driven cavity obeys the equation
\begin{equation} 
a(t+dt) = U_1^\dagger(t,t+dt) U_2^\dagger(t,t+dt) a(t) U_2 (t,t+dt) 
U_1(t,t+dt).
\end{equation}
Note that $U_2(t,t+dt)$ commutes with 
$U_1(t,t+dt)$ because of the finite $\tau$, so the ordering in the above 
equation is not significant. Expanding the terms as above gives 
\begin{eqnarray} 
da &= &\frac{\gamma_1}{2} \left[ (N+1) (2c_1^\dagger a c_1 - a 
c_1^\dagger c_1 - c_1^\dagger c_1 a) + N (2c_1ac_1^\dagger - ac_1 
c_1^\dagger - c_1 c_1^\dagger a) \right. \nonumber \\
&&\;\;\;\;\;\;  \left. + \; M [c_1^\dagger, [ c_1^\dagger,a]] + M^* 
[c_1,[c_1,a]] \right] dt \nonumber \\
&& + \frac{\gamma_2}{2} \left[ (N+1) (2c_2^\dagger a c_2 - a 
c_2^\dagger c_2 - c_2^\dagger c_2 a) + N (2c_2ac_2^\dagger - ac_2 
c_2^\dagger - c_2 c_2^\dagger a) + \right. \nonumber \\
&& \;\;\;\;\;\;  \left. + \; M [c_2^\dagger, [ c_2^\dagger,a]] + M^* 
[c_2,[c_2,a]] \right] dt \nonumber \\
&& -\sqrt{\gamma_1}[dB_1^\dagger c_1 - dB_1 c_1^\dagger, a] - 
\sqrt{\gamma_2} [ dB_2^\dagger  c_2 - dB_2 c_2^\dagger , a] . 
\label{sortofito}
\end{eqnarray}
Here, the implicit time argument of $a$, $c_1$ and $c_2$ is $t$, 
while that of  $dB_2 = dB_1 + \sqrt{\gamma_1} c_1 dt$ is $t-\tau$.

Equation (\ref{sortofito}) is an Ito equation in that it is gives an 
explicit algorithm for calculating an infinitesimal increment in some 
operator, given all of the other operators at the start of the time 
interval. However, it is not an Ito equation in the sense that the noise 
terms are not independent of the other operators. Although the noise 
input $dB_1(t)$ is independent, $dB_1(t-\tau)$ in $dB_2(t-\tau)$ is 
not 
independent of an arbitrary system operator $a(t)$, as can be seen  as 
follows
\begin{equation}
a(t) = a(t-\tau) - \sqrt{\gamma_1}[dB_1^\dagger(t-\tau) c_1(t-\tau) - 
dB_1(t-\tau) c_1^\dagger(t-\tau), a(t-\tau)] + O(\tau). \label{cortau}
\end{equation}
To derive a master equation, it is necessary to take the formal limit 
$\tau \rightarrow 0$. The basic physics of the problem is independent 
of $\tau$ because, as yet, we are not considering feedback. This 
means that corrections of order $\tau$ in Eq.~(\ref{cortau}) can be 
ignored, after calculating the second order Ito corrections. 
 By using this equation carefully, Eq.~(\ref{sortofito}) can be 
converted to an Ito equation in which the noise terms are independent:
\begin{eqnarray} 
da &=& (N+1) \left[ \frac{\gamma_1}{2} (2c_1^\dagger a c_1 - a 
c_1^\dagger c_1 - c_1^\dagger c_1 a) + \frac{\gamma_2}{2} 
(2c_2^\dagger a c_2 - a c_2^\dagger c_2 - c_2^\dagger c_2 a ) 
\right. \nonumber \\
&& \;\;\;\;\;\;\;\;\;\;\;\;\;\;\;\;  \left. +  \sqrt{\gamma_1 \gamma_2} 
(c_2^\dagger a c_1 - a c_2^\dagger c_1 + c_1^\dagger a c_2 - 
c_1^\dagger c_2 a ) \right] \nonumber \\
&& + N \left[ \frac{\gamma_1}{2} (2c_1 a c_1^\dagger - a c_1 
c_1^\dagger - c_1 c_1^\dagger a) + \frac{\gamma_2}{2}( 2c_2 a 
c_2^\dagger - a c_2c_2^\dagger  - c_2 c_2^\dagger a ) 
\right. \nonumber \\
&& \;\;\;\;\;\;\;\; \left. +  \sqrt{\gamma_1 \gamma_2} (c_2 a 
c_1^\dagger - a c_2 c_1^\dagger + c_1 a c_2^\dagger - c_1 
c_2^\dagger a ) \right] \nonumber \\
&& + M \left[ \frac{\gamma_1}{2} [c_1^\dagger,[c_1^\dagger,a]]  + 
\frac{\gamma_2}{2}[c_2^\dagger,[c_2^\dagger,a]] + \sqrt{\gamma_1 
\gamma_2} [c_1^\dagger,[c_2^\dagger,a]]  \right] \nonumber \\
&&+ M^* \left[ \frac{\gamma_1}{2} [c_1,[c_1,a]]  + 
\frac{\gamma_2}{2}[c_2,[c_2,a]] + \sqrt{\gamma_1 \gamma_2} 
[c_1,[c_2,a]] \right] \nonumber \\
&& -\sqrt{\gamma_1}[dB_1^\dagger c_1 - dB_1 c_1^\dagger, a] - 
\sqrt{\gamma_2} [ dB_1^\dagger  c_2 - dB_1 c_2^\dagger , a] . 
\label{trueito}
\end{eqnarray} 
In this equation, all operators have the same time argument. It agrees 
with the results of Refs.~(\cite{Gar93,Car93a}) for the case 
$N=M=0$, but it should be noted that the details of the derivation in 
both of these papers differs from ours.

It is now a simple matter to convert this stochastic Heisenberg 
equation into a master equation for the density operator of both 
systems $W$
\begin{eqnarray}
\dot{W} &=& (N+1) \left[ \gamma_1{\cal D}[c_1] W + 
\gamma_2{\cal D}[c_2] W + \sqrt{\gamma_1\gamma_2}
\left( [c_1 W, c_2^\dagger]  + [c_2, W c_1^\dagger] \right) \right] 
\nonumber \\
&&  + N \left[  \gamma_1{\cal D}[c_1^\dagger] W + \gamma_2{\cal 
D}[c_2^\dagger] W+ \sqrt{\gamma_1\gamma_2}
\left( [c_1^\dagger W, c_2] +[ c_2^\dagger, W c_1]\right) \right] 
\nonumber \\
&& + M \left[ \frac{\gamma_1}{2} [c_1^\dagger,[c_1^\dagger,W]] 
+ \frac{\gamma_2}{2} [c_2^\dagger,[c_2^\dagger,W]] + 
\sqrt{\gamma_1\gamma_2}[c_2^\dagger,[c_1^\dagger,W]] \right] 
\nonumber \\
 &&+ M^* \left[ \frac{\gamma_1}{2} [c_1,[c_1,W]] 
+ \frac{\gamma_2}{2} [c_2,[c_2,W]] + 
\sqrt{\gamma_1\gamma_2}[c_2,[c_1,W]] \right] \nonumber \\
&& + \sqrt{\gamma_1} [ \beta^*(t) c_1 - \beta (t) c_1^\dagger ,W]
 + \sqrt{\gamma_2} [ \beta^*(t) c_2 - \beta (t) c_2^\dagger ,W] - 
i[H,W]  \label{cascsys}
\end{eqnarray}
Here we have finally included the intrinsic evolution for the two 
systems, generated by the Hamiltonian $\hbar H$.
This is the general equation for two open quantum systems, linked 
unidirectionally by a bath of harmonic oscillators with an optical 
frequency coherent amplitude contaminated by white noise.  (In the 
following sections, the coherent amplitude will be ignored.) It is a 
generalization of the equations published by Gardiner \cite{Gar93} 
and Carmichael \cite{Car93a}, which treated only a bath in the 
vacuum state. It has the necessary property that, if $H$ is the sum of 
Hamiltonians operating in the Hilbert subspaces of the two systems, 
then the source system ($c_1$) is unaffected by the driven system 
($c_2$). That is to say, the density operator $\rho$ for the source 
system (obtained by tracing over the driven system) obeys the original 
master equation (\ref{sys1me}). This is evident from the fact that the 
only terms in Eq.~(\ref{cascsys}) containing operators from both 
systems involve an exterior commutator with a driven system 
operator, 
which gives zero when traced over the driven subspace. It is not 
possible in general to derive a 
master equation for the second system alone. Its evolution is literally 
driven by the source system.

\subsection{Feedback}

The simplest imaginable all-optical feedback scheme would be to take 
the light from one end of a cavity and reflect it back into the other, 
using a Faraday isolator to prevent interference from reflections in the 
opposite directions. For 
simplicity, we assume a bath in the vacuum state and a cavity with 
equal transmittivities at both end-mirrors. Denote the input vacuum 
state by $b_1(t) = \nu(t)$, the cavity mode annihilation operator by 
$a$, and the damping rate by $\gamma$. Then the evolution of the 
cavity mode due to the first mirror is
\begin{equation}
\dot{a}(t) = -\frac{\gamma}{2} a(t) - \sqrt{\gamma}\nu(t),
\end{equation}
and the output field is $\sqrt{\gamma}a + \nu(t)$. If the time delay in 
the loop is $\tau$, then the effect of the feedback is
\begin{equation}
\dot{a}_{fb}(t) = -\frac{\gamma}{2} a(t) - 
\sqrt{\gamma}[\sqrt{\gamma} a(t-\tau) + \nu(t-\tau)].
\end{equation}
Provided that $\gamma\tau \ll 1$, the effect of the time delay can be 
ignored, apart from introducing an arbitrary phase factor depending on 
the optical path length of the loop. Because the bath is in a vacuum 
state, it is not necessary to worry about the Ito corrections used in 
going from Eq.~(\ref{sortofito}) to Eq.~(\ref{trueito}). Thus the total 
evolution for $a$ is
\begin{equation}
\dot{a}(t) =  - (1 + e^{i\phi}) [ \gamma a(t) + \sqrt{\gamma} \nu(t) ].
\end{equation}

The master equation equivalent to this quantum Langevin equation is
\begin{equation}
\dot{\rho} = 2\gamma(1+\cos \phi) {\cal D}[a] \rho - i\gamma \sin 
\phi [a^\dagger a ,\rho ].
\end{equation}
Evidently, if $\phi = \pi$, the damping through the mirrors can be 
completely eliminated. Of course, this is an approximation only. In 
reality, the lifetime of the cavity is enhanced by a factor of order 
$1/\gamma\tau \gg 1$. Losses in the loop would also decrease the 
effectiveness of the feedback. For  $\phi = 0$, the damping rate is 
doubled, and for other values of $\phi$, the cavity 
becomes detuned, as well as having its damping rate altered. In 
Sec.~V we will show that these features can be reproduced by 
feedback using a heterodyne detection photocurrent to control a 
driving field at the second mirror. However, electro-optic feedback 
will necessarily introduce extra noise. The all-optical model here 
obviously does not introduce noise; its results are a consequence of 
classical geometrical optics.

\section{Intensity feedback}

In this section we consider feedback of the form described in the 
introduction. The light from the source cavity is used to drive a 
second cavity, which coincides spatially at least in part with the 
source cavity (see Fig.~1). The two modes interact via a nonlinear 
crystal, allowing the driven cavity to control the source cavity. The 
form of the interaction is assumed to depend only on the intensity of 
the driven cavity. It is thus analogous to feedback of the photon flux 
output of the source cavity. We analyze the feedback system using 
both 
master and quantum Langevin equations.

\subsection{Master equation}

The form of our feedback master equation is that of the general 
cascaded systems equation (\ref{cascsys}) derived in the preceding 
section. What distinguishes it as feedback is that the Hamiltonian 
$\hbar H$ is assumed to consist of a source cavity Hamiltonian 
$\hbar H_0$ plus an interaction between the source and driven cavity 
of the form
\begin{equation}
V = \hbar c_2^\dagger c_2 K \label{Vint}
\end{equation}
where $K$ is an Hermitian operator on the source cavity. In this, and 
all following sections, we will measure time in inverse units of the 
decay constant $\gamma_1$ for the source cavity. The decay constant 
$\gamma_2$ for the driven cavity will be denoted $\gamma$. 
Because of the interaction term, the dynamics of the source cavity is 
no longer independent of that of the driven cavity. Thus we must 
consider the master equation for the density operator of both modes 
$W$. According to Eq.~(\ref{cascsys}), this obeys the master 
equation
\begin{equation} \label{intW}
\dot{W} = -i[H_0 + c_2^\dagger c_2 K, W] + {\cal D}[c_1] W+ 
\gamma{\cal D}[c_2]W + \sqrt{\gamma}\left( [c_1 W, c_2^\dagger]  
+ 
[c_2, W c_1^\dagger] \right) .
\end{equation}
Note that here we have assumed that the bath is in the vacuum state 
($N=0$). The reason for this is will become apparent later. 

In deriving Eq.~(\ref{intW}), it is of course necessary to set $\tau=0$. 
However, this is no longer a formal mathematical limit. Rather, it is a 
physical condition that the time delay in the feedback loop should be 
negligible. This is usually desirable in feedback loops. A substantial 
time delay may lead to instabilities and chaos, rather than control. 
Similarly, for the feedback to be effective, the second cavity should 
respond much faster than the first. This ensures that the state of $c_2$ 
is effectively slaved to that of $c_1$, and the  interaction effects 
instantaneous feedback  as far as the source is concerned. Thus, in the 
limit $\gamma \gg 1$, it should be possible to derive a master 
equation including feedback for the source density operator $\rho$ 
alone. To do this we note that since $\gamma \gg 1$ and the bath is in 
the vacuum state, the driven cavity will be very close to being in the 
vacuum state also. This enables the adiabatic elimination procedure 
we 
have used previously \cite{WisMil93a} to be employed again.

We expand $W$ in powers of $1/\sqrt{\gamma}$ as
\begin{equation} \label{Wpow}
W = \rho_0 \otimes |0\rangle\langle 0| + [ \rho_1 \otimes |1\rangle 
\langle 0| + {\rm H.c.} ] + \rho_2 \otimes |1\rangle\langle 1 | ,
\end{equation}
where the $\rho$s exist in the source subspace and the other operators 
in the driven subspace. There is another term of second order in 
$1/\sqrt{\gamma}$, but it plays no part in the procedure which 
follows. Substituting the above expansion into the master equation 
(\ref{intW}) gives the following coupled equations
\begin{mathletters}
\begin{eqnarray}
\dot{\rho}_0 &=& \gamma \rho_2 + \sqrt{\gamma}( \rho_1 
c_1^\dagger + 
c_1\rho_1^\dagger)  + {\cal L}_0 \rho_0  \label{1a} \\ 
\dot{\rho}_1 &=& - \frac{1}{2} \gamma \rho_1 -iK\rho_1 - 
\sqrt{\gamma} [c_1\rho_0 + O(1/\gamma) ]  + {\cal L}_0 \rho_1 
\label{1b} \\
\dot{\rho}_2 &=& -\gamma \rho_2  -i[K,\rho_2] - \sqrt{\gamma}( 
\rho_1 c_1^\dagger + c_1\rho_1^\dagger)  + {\cal L}_0 \rho_2 
,\label{1c}
\end{eqnarray}
\end{mathletters}
Here ${\cal L}_0 \rho \equiv {\cal D}[c_1]\rho -i[H_0,\rho]$. In this 
approximation, the source density operator is $\rho = \rho_0 + \rho_2$ 
which evidently obeys
\begin{equation}
\dot{\rho} = -i[K,\rho_2] + {\cal L}_0 \rho. \label{notme}
\end{equation} 
To turn this into a master equation, we require an expression for 
$\rho_2$ in terms of $\rho \simeq \rho_0$. It is now obvious why we 
assumed a zero temperature bath. For $N$ finite, $\rho_2$ would 
have a finite size irrespective of $\gamma$. Thus the signal due to the 
driving from the source, which is of order $1/\gamma$ would be 
swamped by the noise, and the feedback would not work.

To obtain $\rho_2$ it is first necessary to obtain $\rho_1$. Since 
almost all of the probability is in $\rho_0$, it is evident from 
Eq.~(\ref{1b}) that $\rho_1$ relaxes much rapidly than $\rho_0 
\simeq \rho$. It is thus permissible to set $\rho_1$ equal to its steady 
state value of
\begin{equation}
\rho_1 = \left( 1 + i \frac{2K}{\gamma} \right)^{-1} \frac{-
2}{\sqrt{\gamma}} c_1 \rho. \label{1slav}
\end{equation}
The expression on the right hand side will be a well defined operator 
if we assume that $|K| \ll \gamma$ in some sense. This assumption 
will 
be valid in practice, as single-photon nonlinearities are typically much 
smaller than damping rates. It allows the denominator to be expanded 
to 
first order in $K/\gamma$. Since $\rho_1$ is now slaved to $\rho_0$, 
it will evolve at a rate much smaller than $\gamma$. Thus, from 
Eq.~(\ref{1c}), $\rho_2$ will relax to a steady state determined by the 
slaved value of $\rho_1$:
\begin{equation}
\dot{\rho}_2 =   -\gamma \rho_2  -i[K,\rho_2] + 4 c_1\rho 
c_1^\dagger -4i 
{\gamma}^{-1} [K, c_1 \rho c_1^\dagger ] + {\cal L}_0 \rho_2 .
\end{equation}
The slaved value of $\rho_2$, again to first order in $K/\gamma$, is
\begin{equation}
\rho_2 =  \frac{4 c_1 \rho c_1^\dagger}{\gamma} - 
\frac{4i}{2\gamma} 
\left[ \frac{4K}{\gamma} , c_1 \rho c_1^\dagger \right].
\end{equation}
Substituting this into Eq.~(\ref{notme}) gives the master equation
\begin{equation}
\dot{\rho} = -i[Z,c_1\rho c_1^\dagger] - \frac{1}{2} [Z,[Z,c_1\rho 
c_1^\dagger]] + {\cal D}[c_1]\rho - i [H_0,\rho], \label{box1}
\end{equation}
where we have defined
\begin{equation} Z= 4K / \gamma. \label{defZ} \end{equation}

This master equation is the general equation for Markovian, 
intensity-dependent feedback in the small $Z$ limit. Unfortunately,
 it is not a valid master equation \cite{Gar91}
in the sense that it cannot be written in the form
\begin{equation}
\dot{\rho} = -i[H,\rho] + \sum_\mu {\cal D}[c_\mu] \rho,
\end{equation}
where the $c_\mu$ are arbitrary operators. However, there is an 
equation of this form which is equal to Eq.~(\ref{box1}) when 
expanded to second order in $Z$. That equation is
\begin{equation} \label{box2}
\dot{\rho} = -i[H_0,\rho] + {\cal D}[e^{-iZ} c_1 ] \rho.
\end{equation}
This equation has been previously derived as a feedback master 
equation appropriate to electro-optical feedback \cite{Wis93b}. This 
can be seen more clearly by rewriting Eq.~(\ref{box2}) as
\begin{equation}
\rho(t+dt) = \exp\left[\left( -iH_0 - \frac{1}{2}c_1^\dagger c_1 
\right)dt 
\right] \rho(t) \exp \left[\left( iH_0 - \frac{1}{2}c_1^\dagger 
c_1\right) dt 
\right] + e^{-iZ}c_1 \rho(t) c_1^\dagger e^{iZ} dt .
\end{equation}
The two terms in this equation can be given an interpretation in terms 
of density operators conditioned on possible photon detections. The 
norm of each term gives the probability for the associated event.
The first term  represents the conditioned density operator when no 
photon is detected \cite{WisMil93c}. Note that it is only 
infinitesimally changed, and that the probability for this event is thus 
very close to unity. The second term represents the evolved density 
operator if a photon is detected at the time $t$. This does not happen 
very often, but when it does, the state of the system jumps (changes 
discontinuously) via the application of the operator $c_1$ 
\cite{WisMil93c}. The effect of the feedback is to cause some finite 
unitary evolution immediately following the detection, via the 
operator $e^{-iZ}$. Of course this does not change the norm of this 
term. The physical interpretation of the operator $Z$ is that it can be 
derived from a time-dependent feedback Hamiltonian
\begin{equation}
H_{fb} = \hbar Z I(t), \label{meastHam1}
\end{equation}
where $I(t)$ is the photocurrent derived from the detector, measured 
in 
units of detections per second. This could 
be produced by a variety of electro-optic devices. 

We thus see that there is a strong correspondence between 
electro-optical feedback using direct detection and 
all-optical feedback with the intensity dependent coupling 
(\ref{Vint}).
One conclusion which can be drawn from this analogy is that, as was 
previously established for electro-optic feedback \cite{WisMil93d}, 
the all-optic feedback considered in this section cannot produce a 
nonclassical state in the source cavity if $K$ is a classical operator. 
By ``classical'' in this context we mean  a linear combination of 
$c_1$,$c_1^\dagger$, and $c_1^\dagger c_1$. ($K$ is of course 
Hermitian.) 
Given that the nonlinear interaction Hamiltonian (\ref{Vint}) is 
already of second order in the field of the driven ($c_2$) cavity, any 
coupling with a nonclassical $K$ would require at least a five-wave 
mixing interaction (including an auxiliary pump field). This seems 
exceedingly impractical. Thus we can 
conclude that intensity-dependent all-optical feedback is not a 
practical way to produce a nonclassical source state.

\subsection{Langevin equation}

The above derivations for the all-optical and electro-optical intensity 
feedback were done using a master equation for the density operator. 
It is possible to use the Langevin approach for both types of feedback, 
and this yields some extra information. We begin with the all-optical 
feedback. The quantum Langevin equation equivalent to the master 
equation (\ref{intW}) is
\begin{eqnarray} 
\dot{a} &= &\frac{1}{2}(2c_1^\dagger a c_1 - a c_1^\dagger 
c_1 - c_1^\dagger c_1 a)  + \frac{\gamma}{2}  (2c_2^\dagger a c_2 - 
a c_2^\dagger 
c_2 - c_2^\dagger c_2 a) \nonumber \\
&& -[b_1^\dagger c_1 - b_1c_1^\dagger, a] - 
\sqrt{\gamma} [ b_2^\dagger  c_2 - b_2 c_2^\dagger , a]  + i[H_0 + 
c_2^\dagger c_2 
K, a] .
\end{eqnarray}
Here $b_1$ is the input vacuum field, and $b_2 = b_1 +  c_1$ is the 
output field from the first cavity. This field which is fed back is to be 
understood to be at a slightly earlier time than all of the other 
operators in the above equation. For this reason, it commutes with all 
other operators. If $b_2$ is moved to the rear (far right) of any 
operator expression, the vacuum noise of $b_2$ will not contribute 
to any average, as the annihilation operator will act directly on the 
bath in the vacuum state. Similar remarks hold for moving 
$b_1^\dagger$ 
to the front of any expression. Thus, if we always put the bath 
operators in normal order as described here, then the fact that it is at a 
slightly earlier time can be ignored. Of course, this technique only 
works because the bath is in the vacuum state. In general, the time 
delay causes the corrections derived in Sec.~II. In the remainder of 
this section, we will always put $b_2$ in normal order without 
mentioning this explicitly.

We wish to adiabatically eliminate mode $c_2$. This obeys
\begin{equation}
\dot{c}_2 = - \frac{\gamma}{2} c_2 - \sqrt{\gamma}b_2 - i K c_2.
\end{equation}
Now the relaxation rate $\gamma$ of mode $c_2$ cannot  be much 
greater than the bandwidth of the vacuum fluctuations, which are 
assumed infinite. Hence, it is not strictly possible to slave $c_2$ to the 
vacuum fluctuations in $b_2$. However, as far as mode $c_1$ is 
concerned,  vacuum fluctuations restricted to a bandwidth of 
$\gamma$ are still effectively white, and so their is no harm in 
pretending that $c_2$ is slaved to the original vacuum fluctuations. 
This allows us to write the slaved value of $c_2$ as
\begin{equation}
c_2 = \left[ 1 - \frac{2iK}{\gamma} \right] \frac{-2}{\sqrt{\gamma}} 
b_2
\end{equation}
Substituting this into the equation for a source cavity operator $a$ 
gives
\begin{eqnarray}
\dot{a} &=& \frac{1}{2}(2c_1^\dagger a c_1 - a c_1^\dagger 
c_1 - c_1^\dagger c_1 a)  -\nu ^\dagger[ c_1,a]  +  [c_1^\dagger, a]\nu  
\nonumber \\
&& + i \frac{4}{\gamma} b_2^\dagger [K,a] b_2 - 
\frac{8}{\gamma^2} b_2^\dagger [K,[K,a]] b_2 + i[H_0 , a].
\end{eqnarray}
Here we have put $b_1 = \nu$ because the equation is only valid 
when the input is a vacuum. 
Substituting in the expression for $b_2$ gives
\begin{eqnarray}
\dot{a} &=&  c_1^\dagger a c_1 - \frac{1}{2}a c_1^\dagger 
c_1 - \frac{1}{2} c_1^\dagger c_1 a  -\nu^\dagger[ c_1,a]  
 + [c_1^\dagger, a]\nu  \nonumber \\
&&+ (c_1^\dagger+\nu^\dagger) \left( i[Z,a] - \frac{1}{2}[Z,[Z,a]] 
\right)(c_1+\nu) + i[H_0 , a], \label{lbox1}
\end{eqnarray}
where $Z$ is as in Eq.~(\ref{defZ}). It is easy to verify by tracing 
over 
the bath that this equation is equivalent to the master equation 
(\ref{box1}).

Just as master equation (\ref{box1}) was an approximation to a 
strictly valid master equation, so is the Langevin equation 
(\ref{lbox1}). The details of the following are to be found in 
Ref.~\cite{Wis93b}. The exact Langevin equation can be derived 
from the following Hamiltonian
\begin{equation}
H_{fb} = \hbar Z b_2^\dagger b_2 . \label{nomeastHam1}
\end{equation}
This  turns out to be equivalent to the photocurrent-dependent 
Hamiltonian (\ref{meastHam1}). That is, simply by replacing the 
photocurrent with the output photon flux $b_2^\dagger b_2$, it is 
possible 
to derive a Langevin picture equation equivalent to the full master 
equation (\ref{box2}):
\begin{eqnarray} \label{lbox2}
\dot{a} &=&  c_1^\dagger a c_1 - \frac{1}{2}a c_1^\dagger 
c_1 - \frac{1}{2} c_1^\dagger c_1 a  -\nu^\dagger[ c_1,a]  +  
[c_1^\dagger, a]\nu \nonumber \\
&& + (c_1^\dagger+\nu^\dagger) \left( e^{iZ}ae^{-iZ} - a 
\right)(c_1+\nu) + i[H_0 , a].
\end{eqnarray}
To second order in $Z$, this is equivalent to the all-optical Langevin 
equation derived above. Unlike Eq.~(\ref{lbox1}), however, 
Eq.~(\ref{lbox2})  is a completely valid 
Langevin equation so that if $a_1$ and $a_2$ are two arbitrary 
operators, then the equation of motion it generates for the product 
operator $a_1 a_2$ is equal to what would be obtained from the two 
operators separately, using the Ito rules for quantum stochastic 
calculus
\begin{equation}
d(a_1 a_2) = (da_1) a_2 + a_1(da_2) + (da_1)(da_2).
\end{equation}
Thus it is also possible to find a correspondence between all-optical 
and 
electro-optical feedback using quantum Langevin equations.

One property that a Langevin equation has which the master equation 
lacks is that it simply gives an expression for the field reflected from 
the driven cavity (see Fig.~1). Calling this field $b_3$, we have
\begin{equation}
b_3 = b_2 + \sqrt{\gamma}c_2.
\end{equation}
From the adiabatic expression for $c_2$, we have
\begin{equation}
b_3 =  - \left[1 - \frac{4iK}{\gamma} \right] b_2.
\end{equation}
Apart from a change of sign (due to reflection), this is equal to
what would be obtained 
from the idealized Hamiltonian (\ref{nomeastHam1})
\begin{equation}
b_3 = e^{-iZ}b_2
\end{equation}
to first order in $Z$. The statements which were made above 
regarding the inability of a classical feedback operator $Z$ to produce 
a nonclassical source state do not necessarily apply to the output 
operator $b_3$. It may exhibit nonclassical features even if the source 
state is classical. We will not pursue the properties of this output field 
further in this section, but we will in the next section where we 
consider quadrature-dependent feedback.

\section{Quadrature feedback}

\subsection{Master equation}
We begin this section on quadrature feedback by reproducing the 
calculation of Sec.~III~A, but with the interaction between the source 
and driven cavity 
\begin{equation} \label{quadV}
V = \hbar (c_2 + c_2^\dagger) J ,
\end{equation}
which is linear in the real quadrature of the driven cavity.  Here $J$ is 
an Hermitian operator in the source cavity. In practice, conservation 
of energy would require at least one other auxiliary field which may 
be  treated classically. With the input to the 
source cavity in the vacuum state as before, and with the the damping 
rate $\gamma$ of the second cavity much greater than that of the first 
(set to unity), it is again possible to expand the combined density 
operator in powers of $1/\sqrt{\gamma}$ as in Eq.~(\ref{Wpow}). In 
this case, the source cavity operators obey
\begin{mathletters}
\begin{eqnarray}
\dot{\rho}_0 &=& \gamma \rho_2 + \sqrt{\gamma}( \rho_1 
c_1^\dagger + 
c_1\rho_1^\dagger)  -i[J\rho_1 - \rho_1^\dagger J] + {\cal L}_0 
\rho_0  
\label{2a} \\ 
\dot{\rho}_1 &=& - \frac{1}{2} \gamma \rho_1 -i[J\rho_0 + 
O(1/\gamma)]   - 
\sqrt{\gamma} [c_1\rho_0 + O(1/\gamma) ]  + {\cal L}_0 \rho_1 
\label{2b} \\
\dot{\rho}_2 &=& -\gamma \rho_2  -i[J \rho_1^\dagger - \rho_1 J] - 
\sqrt{\gamma}( \rho_1 c_1^\dagger + c_1\rho_1^\dagger)  + {\cal 
L}_0 \rho_2 .
\label{2c}
\end{eqnarray}
\end{mathletters}
The approximate source density operator $\rho = \rho_0 + \rho_2$ 
 obeys
\begin{equation}
\dot{\rho} = -i[J,\rho_1+\rho_1^\dagger] + {\cal L}_0 \rho. 
\label{notme2}
\end{equation} 

Evidently, to derive a ME for $\rho$ in this case it is only necessary 
for $\rho_1$ to be slaved to $\rho = \rho_0 + O(1/\gamma)$. From 
Eq.~(\ref{2b}), this is obviously true for large $\gamma$, with the 
slaved value
\begin{equation}
\rho_1 = \frac{2}{\gamma} [ -\sqrt{\gamma}c_1 \rho - iJ \rho ].
\end{equation}
Here we are keeping only the lowest order in $1/\gamma$, and 
assuming that $J$ is of order $\sqrt{\gamma}$. Substituting this 
expression into Eq.~(\ref{notme2}) gives the master equation
\begin{equation}
\dot{\rho} = -i[Y,c_1\rho + \rho c_1^\dagger] - \frac{1}{2} 
[Y,[Y,\rho]] + 
{\cal D}[c_1]\rho -i[H_0,\rho], \label{quadme}
\end{equation}
where we have defined
\begin{equation}
Y = \frac{-2J}{\sqrt{\gamma}}, \label{defY}
\end{equation}
which is of order unity. This master equation can be put into the form 
required of master equations thus:
\begin{equation} \label{box3}
\dot{\rho} = {\cal D}[c_1 - iY] \rho -i\left[H_0 + 
\frac{1}{2}(c_1^\dagger 
Y + Yc_1) ,\rho\right].
\end{equation}

This feedback master equation has been previously derived from a 
model of electro-optic feedback using homodyne detection 
\cite{WisMil93b,WisMil93d}. If the output of the cavity is subject to 
unit efficiency homodyne detection of the real quadrature, then the 
state of the cavity conditioned on the measured photocurrent obeys
\begin{equation} \label{homoito1}
d\rho_c(t) = dt \left\{-i[H_0,\rho_c(t)] + {\cal D}[c_1] \rho_c(t) 
\right\} +  dW(t) {\cal H}[c_1]  \rho_c(t),
\end{equation}
where the nonlinear superoperator ${\cal H}$ is defined by
\begin{equation} \label{defH}
{\cal H}[a] \rho = a\rho + \rho a^\dagger - Tr [ (a\rho + \rho 
a^\dagger) ] \rho,
\end{equation}
and where $dW$ is an infinitesimal real Weiner increment satisfying 
$dW(t)^2 = dt$. The subscript $c$ indicates that the state is 
conditioned on the signal photocurrent, which is given by
\begin{equation} \label{homath}
I_c(t) = \langle c_1 + c_1^\dagger \rangle_c(t) + \xi(t) ,
\end{equation}
where $\xi(t)dt  = dW(t)$ represents Gaussian white noise
 \cite{Gar85}. These 
equations can be derived from the quantum jumps mentioned in 
Sec.~III A, in the limit that the local oscillator amplitude used in the 
homodyne detection becomes infinitely large 
\cite{Car93b,WisMil93a}. The feedback is then effected by 
introducing the time-dependent Hamiltonian
\begin{equation} \label{homoham}
H_{fb}(t) =\hbar I_c(t) Y.
\end{equation}
Taking into account that the photocurrent in (\ref{homoham}) is in 
reality smooth [unlike the mathematical expression (\ref{homath})], 
and that its time argument must include a delay, one can derive the 
master equation (\ref{box3}) relatively straightforwardly.

\subsection{Langevin equation}

If the input to the source cavity is not a vacuum, then the technique of 
adiabatic elimination of Sec.~IV~A cannot be used. Unlike the 
intensity dependent feedback of Sec.~III, it is nevertheless possible to 
obtain a sensible result for non-vacuum input using a Langevin 
equation approach. The difference arises because the quadrature flux 
of a broad-band squeezed or thermal state is a well-defined stochastic 
quantity, whereas the photon flux is not well defined (it tends to 
infinity 
in the limit of white noise). Because of the thermal and squeezed 
terms, 
it is necessary to use the  explicit quantum equation 
(\ref{sortofito}) also including the time delay $\tau$. 
Adding the intercavity coupling gives the feedback 
Langevin equation for an arbitrary operator $a$
\begin{eqnarray} 
da &= &\frac{1}{2} \left[ (N+1) (2c_1^\dagger a c_1 - a 
c_1^\dagger c_1 - c_1^\dagger c_1 a) + N (2c_1ac_1^\dagger - ac_1 
c_1^\dagger - c_1 c_1^\dagger a) \right. \nonumber \\
&& \left. \;\;\;\;
+ M [c_1^\dagger, [ c_1^\dagger,a]] + M^* [c_1,[c_1,a]] \right] dt 
\nonumber \\
&& + \frac{\gamma}{2} \left[ (N+1) (2c_2^\dagger a c_2 - a 
c_2^\dagger c_2 - c_2^\dagger c_2 a) + N (2c_2ac_2^\dagger - ac_2 
c_2^\dagger - c_2 c_2^\dagger a) \right. \nonumber \\
&& \left. \;\;\;\;
+ M [c_2^\dagger, [ c_2^\dagger,a]] + M^* [c_2,[c_2,a]] \right] dt 
\nonumber \\
&& -[dB_1^\dagger c_1 - dB_1 c_1^\dagger, a] - 
\sqrt{\gamma} [ dB_2^\dagger  c_2 - dB_2 c_2^\dagger , a],  
+ i[H_0 + (c_2+c_2^\dagger)J,a] dt ,
\label{sortofito2}
\end{eqnarray}
where the time argument of $dB_2 = dB_1 + c_1dt$ is $t-\tau$, 
and all other operators are at time $t$. 
An operator $a_1$ for the source cavity will obey the equation
\begin{eqnarray}
da &= &\frac{1}{2} \left[ (N+1) (2c_1^\dagger a c_1 - a 
c_1^\dagger c_1 - c_1^\dagger c_1 a) + N (2c_1ac_1^\dagger - ac_1 
c_1^\dagger - c_1 c_1^\dagger a)  \right. \nonumber \\
&& \left. \;\;\;\;
+ M [c_1^\dagger, [ c_1^\dagger,a]] + M^* 
[c_1,[c_1,a]] \right] dt \nonumber \\
&& -  [ dB_1^\dagger  c_1 - dB_1 c_1^\dagger , a]  + i[H_0 +  
(c_2+c_2^\dagger)J,a] dt .
\end{eqnarray}

Evidently, to obtain a Langevin equation for a source cavity
operator involving no driven cavity operators, it is necessary
only to adiabatically eliminate $c_2$. From Eq.~(\ref{sortofito2}), 
this obeys 
\begin{equation}
\dot{c_2}(t) = - \frac{\gamma}{2}c_2(t) - \sqrt{\gamma}b_2(t-\tau)  
-iJ(t) .
\end{equation}
As before, if $\gamma \gg 1$, then there is no harm in
replacing $c_2$ by the slaved value
\begin{equation} \label{slavc2}
c_2(t)= \frac{-2}{\sqrt{\gamma}} \left[ b_2(t-\tau) +
\frac{iJ(t)}{\sqrt{\gamma}} \right],
\end{equation}
providing  that the resulting term in the Langevin 
equation is treated in the Stratonovich sense.
That is to say, the effective feedback Hamiltonian
\begin{equation} \label{Heff}
H_{fb}(t) = \hbar (c_2 +c_2^\dagger) J = i\hbar \left[ b_2^\dagger(t-
\tau) 
\left(\frac{2iJ(t)}{\sqrt{\gamma}} \right) - \left(\frac{-
2iJ(t)}{\sqrt{\gamma}}\right) b_2(t-\tau) \right]
\end{equation}
is to be treated in the same manner as the coupling Hamiltonian 
(\ref{H2}). The effective Hamiltonian (\ref{Heff}) has the same 
form as Eq.~(\ref{H2}), but for the replacement of 
$\sqrt{\gamma_2}c_2$ by $-iY$ [as defined above (\ref{defY})]. 
Thus it is possible to use the same procedure to derive a Langevin 
equation for a source cavity operator $a$ in the limit $\tau \to 0$. 
From Eq.~(\ref{trueito}), this is immediately  seen to be
\begin{eqnarray} 
da &=& (N+1) \left[ \frac{1}{2} (2c_1^\dagger a c_1 - a 
c_1^\dagger 
c_1 - c_1^\dagger c_1 a) - \frac{1}{2}[Y,[Y,a]]+ (i[Y, a] c_1  + i 
c_1^\dagger 
[Y,a] ) \right]  \nonumber \\
&& + N \left[ \frac{1}{2} (2c_1 a c_1^\dagger - a c_1 c_1^\dagger - 
c_1 c_1^\dagger a) - \frac{1}{2}[Y,[Y,a]]+ 
 (-i[Y, a] c_1^\dagger -i c_1 [Y,a]  ) \right] \nonumber \\
&& + M \frac{1}{2} \left[ [c_1^\dagger,[c_1^\dagger,a]] -  
[Y,[Y,a]] + 2i[c_1^\dagger,[Y,a]]  \right]  \nonumber \\
&&+ M^* \frac{1}{2} \bigl[ [c_1,[c_1,a]] -  
[Y,[Y,a]] - 2i[c_1,[Y,a]]  \bigr]  \nonumber \\
&& -[dB_1^\dagger (c_1-iY) - dB_1( c_1^\dagger+iY), a] + i[H_0, a 
] dt, 
\label{trueito2}
\end{eqnarray} 
where $dB_1$ is a true Ito increment.

The master equation corresponding to this Langevin equation is now 
easy to derive. It is
\begin{eqnarray} 
\dot{\rho} &=& (N+1) \left\{{\cal D}[c_1] \rho -i[Y, c_1 \rho + \rho 
c_1^\dagger ] \right\} + N \left\{ {\cal D}[c_1^\dagger] \rho + 
i[Y,c_1^\dagger \rho + 
\rho c_1 ] \right\} \nonumber \\
&& + M \left\{ \frac{1}{2}[c_1^\dagger,[c_1^\dagger,\rho]] + 
i[Y,[c_1^\dagger, 
\rho]]  \right\} 
+ M^* \left\{ \frac{1}{2}[c_1,[c_1,\rho]] - i[Y,[c_1, \rho]] \right\} 
\nonumber \\
&& + (2N+1+M+M^*) {\cal D}[Y] \rho -i[H_0,\rho] .
\label{quadmenoise}
\end{eqnarray}
This is the general master equation for quadrature-based feedback in 
the presence of quantum white noise. 
For a vacuum input, it reduces to the master equation derived in the 
preceding section (\ref{box3}). Recall that there are two derivations 
of Eq.~(\ref{box3}), from the all-optical model which we have just 
generalized for a non-vacuum input, and from an electro-optic model 
of feedback based on homodyne detection 
\cite{WisMil93b,WisMil93d}. The correspondence 
between the two models is again close, as they can both be derived 
directly from a time dependent Hamiltonian. In the case of all-optical 
feedback, we have just seen that the effective feedback Hamiltonian 
(\ref{Heff}) is
\begin{equation}\label{Heff2}
H_{fb} = \hbar (b_2 + b_2^\dagger) Y.
\end{equation}
This is identical to the current controlled Hamiltonian 
(\ref{homoham}) if the homodyne signal is identified with the 
quadrature flux output of the cavity.

This close relationship suggests that it should be possible to derive 
the full master equation (\ref{quadmenoise}) from a measurement 
theory approach to homodyne measurement in the presence of noise. 
Such a theory has not been published before, to our knowledge. The 
difficulty in formulating the theory is that any white noise will, in 
theory, cause an infinite photon flux at the detectors. Thus, for any 
finite local oscillator strength, the signal will be swamped, just as it 
is 
for direct detection. A new approach is needed, in which the 
limit of infinite local oscillator strength is taken before determining 
the effect of the measurement on the system. After some effort, it can 
be 
shown that the generalization of the homodyne 
photocurrent expression (\ref{homath}) is 
\begin{equation} \label{genhomath}
I(t) = \langle c_1 + c_1^\dagger \rangle(t)+ \sqrt{L}\, \xi(t),
\end{equation}
where we have defined
\begin{equation}
L = 2N+1+M+M^*, \label{defL}
\end{equation}
and $\xi(t)$ is delta function normalized real Gaussian noise as 
before. Note that $L$ can be less than one (its vacuum value) in the 
case of squeezed input noise. The stochastic evolution of the system 
given this photocurrent is
\begin{eqnarray}
d\rho(t) &=&  dt \left\{(N+1) {\cal D}[c_1]\rho + N 
{\cal D}[c_1^\dagger]\rho  + \frac{M}{2} 
[c_1^\dagger,[c_1^\dagger,\rho]] + 
\frac{M^*}{2} [c_1,[c_1,\rho]] - i[H_0,\rho] \right\} \nonumber \\
&&+ \frac{1}{\sqrt{L}}\, dW(t) \, {\cal H} \left[ (N+M+1) c - 
(N+M^*) c^\dagger \right]  \rho,
\end{eqnarray}
where the nonlinear superoperator ${\cal H}$ is as defined in 
(\ref{defH}). If the current (\ref{genhomath}) is fed back in the same 
manner as before [Eq.~(\ref{homoham})], then a master equation can 
be derived using the same procedure as in the case of a vacuum input. 
The result is, not surprisingly, Eq.~(\ref{quadmenoise}). Thus, there 
is a complete correspondence between electro-optic and all-optical 
feedback in the case of quadrature-dependent feedback. 

\subsection{In-loop and output squeezing}

As with intensity-dependent feedback, the all-optical quadrature-% 
dependent feedback scheme described above is not a good way to try 
to produce nonclassical light in the source cavity. This is evident from 
the master equation (\ref{quadme}). Ignoring $H_0$, this will only 
produce squeezing if $Y$ is a nonclassical operator  (see preceding 
section). Given the origin of $Y$ in the original coupling 
(\ref{quadV}), it is evident that at least a $\chi^{(3)}$ nonlinearity 
would be required to produce nonclassical source light. However, as 
noted in the preceding section, this requirement does not hold for the 
production of squeezing in the output light from the second cavity. Of 
course, this beam of light does not exist in the electro-optic case; it is 
only present for an all-optical feedback system. It turns out that this 
light can exhibit complete squeezing on-resonance, even though the 
source cavity is in a classical state.

It is easiest to examine the in-loop and output fields in the Langevin 
equation approach. The output field operator $b_3$ from the driven 
cavity is related to the in-loop field (denoted $b_2$ as above) by
\begin{equation}
b_3 = b_2 + \sqrt{\gamma} c_2.
\end{equation}
Using the slaved value of $c_2$ (\ref{slavc2}), this gives
\begin{equation}
b_3 = -(b_2 - iY) = - (b_1+c_1-iY). \label{b3quad}
\end{equation}
Apart from the unimportant sign change, this is just what would be 
expected from the effective Hamiltonian (\ref{Heff2}).

Now consider the simplest possible case, where $b_1$ is in a vacuum 
state and $Y$ is proportional to the imaginary quadrature of the 
source cavity
\begin{equation}
Y = -\frac{\lambda}{2}(-ic_1 + ic_1^\dagger).
\end{equation} 
This gives linear equations of motion by which the feedback alters the 
statistics of the real quadrature. Specifically, using 
$x=c_1+c_1^\dagger$ 
and $y=-ic_1 + ic_1^\dagger$, we get from Eq.~(\ref{trueito2})
\begin{eqnarray}
\dot{x} &=& -\left( \frac{1}{2} x + \nu_x \right) - \lambda (x + 
\nu_x ) \label{dotx} \\
\dot{y} &=& -\left( \frac{1}{2} y + \nu_y \right) . \label{doty}
\end{eqnarray}
Here $\nu_x$ and $\nu_y$ are vacuum quadrature noise operators 
which obey 
\begin{eqnarray}
\langle \nu_x (t) \nu_x(t') \rangle = \langle \nu_y (t) \nu_y(t')
\rangle  &=& \delta(t-t') \label{nuxyav} \\
 \left[\nu_x(t),\nu_y(t') \right] &=& 2i \delta(t-t'). \label{nuxycr}
\end{eqnarray}
Since the equations for the two quadratures are uncoupled, the 
operator nature of $\nu_x$ and $\nu_y$ (as evidenced by 
Eq.~(\ref{nuxycr}) is mostly unimportant. Hence it is easy to derive 
the steady state variance in $x$ to be
\begin{equation} \label{Vxssq}
V(x)_{ss} = \frac{(1+\lambda)^2}{1+2\lambda}.
\end{equation}
This is always greater than one for any non-zero $\lambda$, which is 
as expected since this sort of feedback cannot produce a squeezed 
intracavity state. 

To examine the extracavity fields, it is necessary to consider the noise 
at different frequencies. Defining the Fourier transform of a quantity 
by
\begin{equation} \tilde{a}(\omega) = \int_{-\infty}^{\infty} 
e^{i\omega t} a(t) dt ,
\end{equation}
Eqs.~(\ref{dotx},\ref{doty}) become
\begin{eqnarray}
\tilde{x}(\omega) &=& -\frac{1 + \lambda}{\frac{1}{2} + \lambda 
-i\omega } \tilde{\nu}_x(\omega) \\
\tilde{y}(\omega) &=& -\frac{1}{\frac{1}{2} - i\omega} 
\tilde{\nu}_y(\omega) .
\end{eqnarray}
The frequency domain counterparts to the time domain relationships 
(\ref{nuxyav},\ref{nuxycr}) are identical but for the replacement of 
$\delta(t-t')$ by $2\pi \delta(\omega + \omega')$. The in-loop 
quadratures of $b_2$ are
\begin{eqnarray}
\tilde{x}_2(\omega) = &=& -\frac{\frac{1}{2} + 
i\omega}{\frac{1}{2} + \lambda -i\omega } \tilde{\nu}_x(\omega) \\
\tilde{y}_2(\omega) &=& -\frac{\frac{1}{2} + 
i\omega}{\frac{1}{2} - i\omega } \tilde{\nu}_y(\omega).
\end{eqnarray}
These obey the commutation relations
\begin{equation} \label{ilcr}
[\tilde{x}_2(\omega),\tilde{y}_2(\omega') ] = 4 \pi i
\delta(\omega+\omega') \times \frac{\frac{1}{4} + 
\omega^2}{\frac{1}{4} + \omega^2 + \lambda \left( \frac{1}{2} + 
i\omega \right) }.
\end{equation}

At first sight, the presence of the second factor here would seem to be 
a flaw in the theory, as it shows that the in-loop field does not obey 
the usual commutation relations for a free field. The commutator in 
Eq.~(\ref{ilcr}) vanishes for low frequencies in the limit $\lambda 
\rightarrow \infty$. However, it must be remembered that the 
canonical commutation relations (\ref{ccr}) for the electromagnetic 
field are defined 
between the field at different points in space, but at the same time. If 
they are defined in terms of frequency components, as here, then these 
are strictly speaking spatial, not temporal, frequencies. It is only for a 
field which is free to propagate over an infinite space that the 
distinction vanishes. In our case, the in-loop field is spatially confined 
to a length $c\tau$, and we have let $\tau$ go to $0$. If we put $\tau$ 
back in the equations, then it is not difficult to see that the only 
modification is to replace $\lambda $ in the frequency domain by 
$\lambda e^{i\omega \tau}$. With this replacement in the above 
frequency commutation relations, it is possible to show \cite{Sha87} 
that the time commutation relations are only changed for times greater 
than $\tau$. That is to say, the canonical (spatial) commutation 
relations are never violated. The field with operator $b_2(t)$ does not 
persist for longer than the time $\tau$ as it travels from the first 
to the second cavity, and so there is never in existence two values of 
the field which violate the canonical commutation relations.

Returning to the output field, we find from Eq.~(\ref{b3quad}) that 
the quadratures of the output field $b_3$ are (including the time delay 
$\tau$), 
\begin{eqnarray}
\tilde{x}_3(\omega) &=&  \frac{e^{i\omega \tau}\left(\frac{1}{2} 
+ i\omega\right)}{\frac{1}{2} + \lambda e^{i \omega \tau} -i\omega 
} \tilde{\nu}_x(\omega) \\
\tilde{y}_3(\omega) &=& \frac{e^{i\omega\tau} \left(\frac{1}{2} +  
i\omega \right) + \lambda}{\frac{1}{2} - i\omega } 
\tilde{\nu}_y(\omega).
\end{eqnarray}
That is to say, the $x$ quadrature of the output is unchanged from the 
$b_2$ value (apart from the phase change due to the time delay), but 
the $y$ quadrature has picked up an extra term. This extra term 
ensures that 
\begin{equation}
[\tilde{x}_3(\omega),\tilde{y}_3(\omega') ] = 4\pi i
\delta(\omega+\omega') ,
\end{equation}
as required because these fields may propagate to infinity. The 
spectrum for the real quadrature (which is equal to the spectrum of the 
photocurrent from a homodyne measurement) is defined by
\begin{equation}
S^x (\omega) = \int_{-\infty}^{\infty} \langle \tilde{x}(\omega) 
\tilde{x}(-\omega') \rangle d\omega',
\end{equation}
and similarly for $y$. For the output field $b_3$ we get
\begin{eqnarray}
S^x_3(\omega) &=& \frac{ \frac{1}{4} + \omega^2 }{| 
\frac{1}{2} + \lambda e^{i\omega \tau} -i\omega |^2 } \\
S^y_3(\omega) &=& 1/S^x_3(\omega) .
\end{eqnarray}
Note that for $\lambda \rightarrow \infty$, we get perfect squeezing at 
low 
frequencies for the $x$ quadrature, and infinite noise in the $y$ 
quadrature, as required by Heisenberg's uncertainty relations.

Is there a simple way to understand this result, that the output may be 
perfectly squeezed, even though the cavity variance in $x$ is 
unbounded [see Eq.~(\ref{Vxssq})]? We look for an answer by an 
analogy with electro-optic feedback. Obviously the above analogy 
with electro-optic feedback mediated by homodyne detection will not 
suffice, because such detection destroys the output beam. 
What would be necessary would be a QND (quantum non-demolition) 
\cite{AlsMilWal88} measurement of the output field. Feedback based 
on a QND measurement of output quadrature would have the same 
effect 
on the intracavity field as homodyne detection, provided it was 
efficient. It is not able to produce intracavity squeezing because an 
extracavity measurement is a poor measurement of the intracavity 
quadrature. However, if the output field $x$ quadrature emerges from 
the QND device unchanged (or relatively little changed), then this 
quantity (the output field quadrature) 
can be well controlled by feedback. The fact that the feedback acts on 
the source cavity is relevant only so far as it affects the output. Such 
QND feedback schemes have been considered by Yamamoto {\em et 
al} \cite{YamImoMac86,HauYam86} and Shapiro {\em et al} 
\cite{Sha87}.

 In the all-optical quadrature feedback scheme, the second cavity can 
be considered as a QND apparatus for the $x$ quadrature output of 
the source cavity. As shown 
above, it does not alter the statistics of the $x$ quadrature which 
reflects from it. However, it increases the variance of the output $y$ 
quadrature, as required for a QND apparatus. The nonlinearity used in 
this case (coupling the quadrature of one mode to that of another) is 
precisely what has been used to model ideal QND quadrature 
measurements \cite{AlsMilWal88,WisMil93d}. Rather than giving a 
current out as its measurement result, the output is directly coupled 
into the dynamics of the source cavity via the interaction Hamiltonian 
(\ref{quadV}). This enables fluctuations in the quantity it is 
monitoring (the output $x$ quadrature) to be suppressed to an 
arbitrary degree. The Langevin equations for the all-optical case are 
identical to those of the 
electro-optic case, once the intervening current in the latter case has 
been eliminated. Thus we see that the correspondence between all-%
optical and electro-optic feedback can be made complete by using 
QND detectors, rather than normal (quantum demolition) 
photodetectors. 

\section{Complex amplitude feedback}

The two all-optical feedback schemes analyzed so far have both had 
an equivalent electro-optic scheme. The reason for this is that the 
couplings between driven and source cavity were QND couplings for 
the driven cavity. Since the driven cavity is slaved to the output of the 
source cavity, that means that the coupling effectively acts as a 
measurement of the output field of the source cavity, the result of 
which directly acts on the source cavity. In this section, we consider a 
Hamiltonian coupling between the two cavities which does not 
factorize as the direct product of a Hermitian operator in each cavity. 
We shall see that the feedback master equation obtained cannot be 
derived from any electro-optic scheme.

\subsection{General equations}

Consider the following Hamiltonian coupling between the two 
cavities
\begin{equation}
V = \hbar (c_2 B^\dagger + c_2^\dagger B),
\end{equation}
where $B$ is an operator on the source cavity. If $B$ is Hermitian (up 
to a phase factor), then this coupling is equivalent to that considered 
in the previous section on quadrature-dependent feedback. In general, 
however, this feedback is sensitive to both quadratures simultaneously 
and hence we have dubbed it complex amplitude feedback. 
Nevertheless, the analysis is identical to that used in the case of 
quadrature feedback. We will quote the main results.

The master equation arising from complex amplitude feedback, 
including a squeezed or thermal input bath to the source cavity, is
\begin{eqnarray}
\dot{\rho} &=& (N+1) \left\{{\cal D}[c_1+A] \rho 
-i\left[ \case{i}/{2}(c_1^\dagger A - A^\dagger c_1) , \rho \right]   
\right\} + N \left\{ {\cal D}[c_1^\dagger+ A^\dagger] \rho - 
i\left[\case{i}{2}( c_1 A^\dagger - A c_1^\dagger ),\rho \right]  
\right\} \nonumber \\
&& + M \left\{ \frac{1}{2}[c_1^\dagger+A^\dagger, 
[c_1^\dagger+A^\dagger,\rho]]  + 
i\left[\case{i}{2}[ c_1^\dagger, A^\dagger] ,\rho \right]  \right\}  
\nonumber \\
&&+ M^* \left\{ \frac{1}{2}[c_1+A,[c_1+A,\rho]]  + 
i\left[\case{i}{2}[ c_1, A] ,\rho \right] \right\}  -i[H_0,\rho]  
\label{came}
\end{eqnarray}
where
\begin{equation}
A =  \frac{2iB}{\sqrt{\gamma}},
\end{equation}
where $\gamma$ is the large damping rate of the driven cavity as 
previously. It can be verified that in the case $A = -iY$, where $Y$ is 
Hermitian, this equation is equal to Eq.~(\ref{quadmenoise}) for 
quadrature feedback. 

One feature which distinguishes Eq.~(\ref{came}) from the 
quadrature feedback equation (\ref{quadmenoise}) is that it can 
produce a nonclassical state in the source cavity even if $A$ is linear 
in $c_1$ and $c_1^\dagger$. To see this,  consider the case $N=0$, 
$H_0=0$ to prevent any obscuring effects. The feedback master 
equation can be rewritten as
\begin{equation}
\dot{\rho} = {\cal D}[c_1]\rho + {\cal D}[A] + (A\rho c_1^\dagger + 
c_1\rho A^\dagger - A^\dagger c_1\rho - \rho c_1^\dagger A).
\end{equation}
For $A$ linear, this equation can be converted into a Fokker-Planck 
equation for the Glauber-Sudarshan $P$ function representation of 
the density operator \cite{Gar91}. The condition for an initially
positive $P$  function (representing a classical state) to remain so is
that the  diffusion matrix be positive semi definite.
The first term (damping) and third term (enclosed in round brackets 
above) will only give first order derivatives. Thus we need consider 
only the second term. Let  
\begin{equation} \label{linA}
A = \frac{\lambda}{2}(c_1+\mu c_1^\dagger).
 \end{equation}
Then it can be readily shown that the eigenvalues of the diffusion 
matrix are proportional to $|\mu|^2 \pm |\mu|$. That is to say, if $0< 
|\mu| < 1$, then this complex amplitude feedback will produce a 
nonclassical state in the cavity. The $|\mu|=1$ limit gives the case of 
quadrature feedback, which, as we showed in Sec.~IV C, 
cannot produce a nonclassical intracavity state. The other limit at 
$|\mu|=0$ corresponds to the simple feedback considered in Sec.~II 
B, which does not even require a nonlinear crystal (it is a classical 
geometrical optics problem).  The property of nonclassicality 
distinguishes this 
all-optical feedback from any form of electro-optic feedback, because 
the latter is known not to produce nonclassical light from a linear 
feedback operator.

 The effective feedback operator (\ref{linA}) can be achieved from 
the interaction Hamiltonian
\begin{equation}
V = -i \hbar g \left[ (c_2 c_1^\dagger 
+ c_2^\dagger c_1) + \mu( c_2 c_1 + c_2^\dagger c_1^\dagger ) 
\right] , \label{realV}
\end{equation}
where we have put $\lambda$ and $\mu$ real for simplicity, and 
where $g=\sqrt{\gamma} \lambda/4$. Let the second mode $c_2$ be 
an orthogonal  polarization mode of the cavity containing the first 
mode $c_1$. That is to say, the second cavity is physically the same 
as the first, unlike the diagrammatic representation in Fig.~1. Then the 
first term in the Hamiltonian (\ref{realV}) could describe mode 
conversion, via a polarization rotator. The coupling constant $g$ 
would be proportional to the (small) proportion of light converted at 
each pass, divided by the round trip time of the cavity. The second 
term, with strength $g\mu$, could only be produced by a nonlinear 
medium, such as a $\chi^{(2)}$ crystal. The two polarization modes 
would be the signal and idler, and the second harmonic would have to 
be strongly driven and heavily damped so that it could be 
adiabatically eliminated. There are no obvious bars to setting up this 
scheme experimentally. 

The Heisenberg picture Ito equation equivalent to the master equation 
(\ref{came}) is
\begin{eqnarray}
da &=& \frac{(N+1)}{2} \left(  2 \bar{c}_1^\dagger a \bar{c}_1 - 
a\bar{c}_1^\dagger \bar{c}_1 - \bar{c}_1^\dagger \bar{c}_1 a - 
[c_1^\dagger A - A^\dagger c_1,a] \right)dt   \nonumber \\
&& + \frac{N}{2}  \left(  2 \bar{c}_1 a 
\bar{c}_1^\dagger - a\bar{c}_1 \bar{c}_1^\dagger - \bar{c}_1 
\bar{c}_1^\dagger a - [c_1 A^\dagger - A c_1^\dagger,a] \right)dt  
\nonumber \\
&&+ \frac{M}{2}  \left(  [\bar{c}_1^\dagger,[\bar{c}_1^\dagger,a]]  
+ [[c_1^\dagger,A^\dagger],a] \right)dt +  \frac{M^*}{2}  \left(  
[\bar{c}_1,[\bar{c}_1,a]]  + [[c_1,A],a] \right)dt \nonumber \\
&& - [dB_1^\dagger \bar{c}_1 - dB_1 \bar{c}_1^\dagger ,a ] + 
i[H_0,a]dt, \label{newrf}
\end{eqnarray}
where we have defined
\begin{equation}
\bar{c}_1 = c_1 + A.
\end{equation}
This equation can be derived from the effective feedback Hamiltonian
\begin{equation} H_{fb} = i\hbar ( b_2^\dagger A - A^\dagger b_2 ). 
\label{cafbH}
\end{equation}
The output field, reflected  off the mirror of the second cavity, is 
\begin{equation} b_3 = -(b_2 + A) = - (b_1 + c_1 + A).
\end{equation}
Again, these equations are equivalent to the quadrature feedback
equations when $A=-iY$.

Consider the case where $A$ is given by Eq.~(\ref{linA}), with 
$\lambda$ real and positive and $\mu$ real.  This gives independent 
linear equations for the quadratures $x,y$ of the field defined in 
Sec.~IV C. Specifically, from Eq.~(\ref{newrf}) with $N=M=0$, one 
obtains
\begin{equation}
\dot{x} = -\frac{1}{2} \left[ 1+ \lambda(1-\mu)+ 
\left(\frac{\lambda}{2}\right)^2(1-\mu^2) \right] x
-\left[1+ \frac{\lambda}{2}(1-\mu) \right]\nu_x,
\end{equation}
where $\nu_x$ is as defined in Sec.~IV C. The equation for the $y$ 
quadrature is identical, but for the replacement of $x$ by $y$, 
$\nu_x$ by $\nu_y$, and $\mu$ by $-\mu$. Note that for $\mu=-1$, 
these equations agree with Eqs.~(\ref{dotx},\ref{doty}), as required.
The intracavity steady state variance for $x$ is 
\begin{equation} \label{Vxssq2}
V(x)_{ss} = \frac{1+\lambda(1-
\mu)+\left(\frac{\lambda}{2}\right)^2(1-\mu)^2}{1+\lambda(1-
\mu)+\left(\frac{\lambda}{2}\right)^2(1-\mu^2)}.
\end{equation}
Note that for $0<\mu<1$, this implies a variance in $x$ less than the 
unit variance of a coherent state. For $0<-\mu<1$, the variance in $x$ 
will be greater than one, but that for $y$ [obtained by replacing $\mu$ 
by $-\mu$ in Eq.~(\ref{Vxssq2})] will be less than one. This is in 
accord with the result stated above, that for $0<|\mu| < 1$, this linear 
all-optical feedback will produce a non-classical intracavity state.
Furthermore, in the limit where $\lambda$ is very large, and 
$\epsilon=1-\mu$ is very small (but not as small as $\lambda^{-1}$), 
then
\begin{equation}
V(x)_{ss} \to \epsilon/2.
\end{equation}
That is to say, the intracavity state can be  arbitrarily squeezed. This 
ideal result could preumably be obtained from electro-optic feedback, 
with some form of nonlinear feedback Hamiltonian. However the 
nonlinear nature of such feedback is quite different from the linear 
all-optical feedback considered here. 

Now consider the output squeezing. From the method of Sec.~IV C, 
the output $x$ spectrum can be calculated to be
\begin{equation}
S^x_3(\omega) = \frac{ \frac{1}{4}(\sigma+\lambda\mu)^2 + 
\omega^2 }{\frac{1}{4}(\sigma - \lambda\mu)^2 + \omega^2 } ,
\end{equation}
where
\begin{equation}
\sigma = 1+\lambda+\left(\frac{\lambda}{2}\right)^2(1-\mu^2).
\end{equation}
The spectrum $S^y_3(\omega)$ can be found by replacing $\mu$ by 
$-\mu$, as before. As shown in Sec.~IV C, the output of the feedback 
loop can show perfect squeezing even for `classical' feedback with 
$|\mu|=1$. Consider $0<|\mu|<1$, so that $\sigma$ is always positive. 
For low frequencies, the output $x$ quadrature is squeezed, with 
$S^x_3(0)<1$, for $\mu<0$. Note that this is the sign of $\mu$ which 
produces intracavity squeezing in the $y$ quadrature. For $\mu>0$, 
the output $y$ quadrature is squeezed, while inside the cavity the $x$ 
quadrature exhibits the nonclassical statistics. In understanding these 
counter-intuitive results, it must be remembered that the output beam 
$b_3$ is not simply the output of the source cavity; the statistics of 
both quadratures are changed by its action as the feedback control 
beam. 

\subsection{Electro-optic analog}

As stated above, all-optical complex amplitude feedback has no 
electro-optical counterpart in general. This is because it is not possible 
to measure both the real and imaginary quadratures of the output field 
simultaneously with unit efficiency. Even a QND measurement of one 
quadrature would introduce noise into the other and so prevent a 
measurement of both. However, it is possible to do two inefficient 
measurements of both quadrature with the two efficiencies adding to 
one (or less than one in practice). It is simplest to consider heterodyne 
detection, which is equivalent to a homodyne measurement of each 
quadrature, each with efficiency of one half. The effect of inefficient 
measurement is to increase the noise introduced by the feedback by a 
factor inversely proportional to the efficiency. The precise meaning of 
this statement will become clear shortly.

In order to compare this  all-optical complex amplitude feedback to 
feedback from heterodyne detection, it is convenient to rewrite 
Eq.~(\ref{came}) in terms of the Hermitian operators $X,Y$ defined 
by
\begin{equation}
A=X-iY.
\end{equation}
We obtain
\begin{eqnarray}
\dot{\rho} &=& -i[H_0,\rho] + (N+1) \left\{{\cal D}[c_1]\rho 
-i[Y,c_1\rho+\rho c_1^\dagger] -i[X,-ic_1\rho+i\rho c_1^\dagger] + 
{\cal D}[X-iY]\rho \right\} \nonumber \\
&&+  N \left\{ {\cal D}[c_1^\dagger] \rho + i[Y,c_1^\dagger \rho + 
\rho c_1 ] -i[X,ic_1^\dagger \rho - i\rho c_1] + {\cal D}[X+iY]\rho 
\right\} \nonumber \\
&& + M \left\{ \frac{1}{2}[c_1^\dagger,[c_1^\dagger,\rho]]  
+i[Y,[c_1^\dagger , \rho]] -i[X,[ic_1^\dagger,\rho]]  - \frac{1}{2} 
[X+iY,[X+iY,\rho]] \right\} \nonumber \\
&& + M^* \left\{ \frac{1}{2}[c_1,[c_1,\rho]]  -i[Y,[c_1,\rho]]  
+i[X,[-ic_1,\rho]] + \frac{1}{2} [X-iY,[X-iY,\rho]] \right\}  
. \label{came2}
\end{eqnarray}
If $A$ is linear in the field amplitude, then $X$ and $Y$ are 
proportional to two orthogonal quadratures of the field. For
controlling noise, it would be sensible for these to be proportional to
the $x$ and $y$ quadratures respectively.

The terms in this equation linear in the feedback operators $X$ and 
$Y$ are reminiscent of the terms in the quadrature dependent 
feedback master equation (\ref{quadmenoise}). However, the terms 
bilinear in $X, Y$, which we will refer to as noise terms for reasons 
which will become evident later, are different. As noted in the 
preceding section, quadrature feedback can be effected by either an 
all-optical scheme or a homodyne detection scheme. In the latter case, 
the master equation was derived from the stochastic evolution 
equation describing the effect of homodyne detection on the system. 
The equivalent equation for heterodyne detection is found to be
\begin{eqnarray}
d\rho(t) &=&  dt \left\{  (N+1) {\cal D}[c_1]\rho + N 
{\cal D}[c_1^\dagger]\rho  + \frac{M}{2} 
[c_1^\dagger,[c_1^\dagger,\rho]] + 
\frac{M^*}{2} [c_1,[c_1,\rho]] - i[H_0,\rho] \right\} \nonumber \\
&&+ \frac{1}{\sqrt{2L_x}}\, dW_x(t) \, {\cal H} \left[ (N+M+1) c - 
(N+M^*) c^\dagger \right]  \rho \nonumber \\
&& + \frac{1}{\sqrt{2L_y}}\, dW_y(t) \, 
{\cal H} \left[ (N-M+1)(-i c) - (N-M^*) (ic^\dagger) \right]  \rho,
\label{heterostoch} \end{eqnarray}
where
\begin{eqnarray}
L_x &=& 2N + 1 + M + M^* \\
L_y &=& 2N + 1 - M - M^*.
\end{eqnarray}
Here, $dW_x(t)$ and $dW_y(t)$ are real infinitesimal Weiner
increments. Physically, they are related to the noise $\xi(t)$ in the two
photocurrents which arise from the heterodyne detection by $dW_q(t) 
=\xi_q(t) dt$ ($q=x,y$). These currents are given by
\begin{eqnarray}
I_x(t) &=& \langle c_1 + c_1^\dagger \rangle + \sqrt{2L_x}\, \xi_x(t) 
\\
I_y(t) &=& \langle -ic_1 + ic_1^\dagger \rangle + \sqrt{2L_y} 
\,\xi_y(t),
\end{eqnarray}
where the normalization has been chosen to make the deterministic 
term the same as in the homodyne case. Note that the noise is greater 
in this case, with $L_q$ multiplied by 2. This is because
each measurement is of efficiency half. In general, if the efficiency of 
the extra-cavity measurement of the quadrature $q$ are $\eta_q$ (such 
that $\eta_x+\eta_y \le 1$), then $2L_q$ is replaced by $L_q/\eta_q$ 
in the above equations. 

Now the stochastic equation (\ref{heterostoch}) can be used to derive 
a feedback master equation analogous to the all-optical
Eq.~(\ref{came2}). The currents control the time-dependent feedback
Hamiltonian
\begin{equation}
H_{fb}(t) = \hbar [ I_x(t) Y + I_y(t) X ]. \label{heterofbH}
\end{equation}
This should be compared with Eq.~(\ref{cafbH}). It is not equivalent 
to
that equation, because there is more noise in the currents than there
is in the quadratures. If one wished to represent the currents by
operators as was done for the case of homodyne detection, then this 
extra noise would enter at the beam splitter necessary to split the
output into two separate homodyne devices. By adding the
evolution from the Hamiltonian (\ref{heterofbH}) to that of
Eq.~(\ref{heterostoch}), we obtain the heterodyne feedback master
equation
\begin{eqnarray}
\dot{\rho} &=& (N+1) \left\{{\cal D}[c_1]\rho -i[Y,c_1\rho+\rho 
c_1^\dagger] -i[X,-ic_1\rho+i\rho c_1^\dagger] \right\} \nonumber \\
&& +  N \left\{ {\cal D}[c_1^\dagger] \rho + i[Y,c_1^\dagger \rho + 
\rho c_1 ] -i[X,ic_1^\dagger \rho - i\rho c_1] \right\} \nonumber \\
&& + M \left\{ \frac{1}{2}[c_1^\dagger,[c_1^\dagger,\rho]]  
+i[Y,[c_1^\dagger , \rho] ] -i[X,[ic_1^\dagger,\rho]] \right\} 
\nonumber \\
&& + M^* \left\{ \frac{1}{2}[c_1,[c_1,\rho]]  -i[Y,[c_1,\rho]]  
+i[X,[-ic_1,\rho]] \right\}  \nonumber \\
&& + 2L_x {\cal D}[Y]\rho + 2L_y{\cal D}[X]\rho -i[H_0,\rho].
\label{heterome2} 
\end{eqnarray}
Note that the desired feedback terms (linear in $X$ and $Y$) are the
same in this equation as in the all-optical Eq.~(\ref{came2}), but the
diffusion terms are different.

To elucidate this difference, we return to the most basic 
example of all-optical feedback, considered in Sec.~II B. The output 
of the source cavity is simply fed back into another mirror of that
cavity. Such feedback is covered by the master equation derived in 
this section, with 
\begin{equation} A = \sqrt{\gamma} e^{i\phi} a.
\end{equation}
Here, $a$ is the annihilation operator for the source cavity with decay
rate $\gamma$. In this simple case, there is no need for a second
cavity. As derived in Sec.~II, the master equation for the cavity is
\begin{equation}
\dot{\rho} = 2\gamma(1+\cos \phi) {\cal D}[a] \rho - i\gamma \sin 
\phi [a^\dagger a ,\rho ], \label{came3}
\end{equation}
where the input is in the vacuum state.
Attempting to replicate this feedback by using heterodyne
detection yields the master equation 
\begin{equation}
\dot{\rho} = 2\gamma(1+\cos \phi) {\cal D}[a] \rho - i\gamma \sin 
\phi [a^\dagger a ,\rho ] + \gamma( {\cal D}[a] + {\cal D}[a^\dagger] 
)\rho .
\end{equation}
The equation of motion for the mean field from this equation is
identical to that of Eq.~(\ref{came3}). However, the presence of the
extra term introduces noise into both quadratures equally. If
$\phi=\pi$, so that the deterministic dynamics are eliminated, then the
variance in each quadrature will simply grow linearly. This clearly
shows the effect of the noise introduced by attempting to measure  
both quadratures in electro-optic feedback, as opposed to the coherent
back coupling of both quadratures in all-optical feedback.

\section{Conclusion}

At the simplest level, feedback is a process by which a system 
influences itself through its action on a second system. Usually, the 
second system would be a complicated feedback apparatus, 
consisting of measurement devices, signal processors and the like. 
However, it is always possible to conceive of more direct schemes, in 
which the feedback loop is treated on the same level as the system. In 
this paper we have examined feedback on optical cavities. The usual 
implementation of optical feedback is electro-optic feedback. The 
light emitted by the cavity enters a detector, and the photocurrent 
produced is used to control the dynamics of the cavity by some 
electro-optic devices. The more direct method is all-optical feedback. 
We have considered turning the output beam from the source cavity 
onto a second cavity which is directly coupled to the first by some 
nonlinear crystal. To compare these two methods at the quantum limit, 
we assumed that the feedback could be approximated as a Markovian 
process. In the electro-optic case, this corresponds to assuming that 
the overall time delay in the feedback loop is much smaller than the 
lifetime of the source cavity, while in the all-optical case, we also 
need the linewidth of the second cavity to be much greater than that of 
the first. Under these conditions, a master equation, or quantum 
Langevin equation can be derived for the source cavity alone.

Our main conclusion is that there is a strong relationship between all 
optical and electro-optical quantum-limited feedback. In fact, if the 
direct coupling Hamiltonian $V$ between the two cavities factorizes 
as the product of Hermitian operators in each cavity, then the all 
optical feedback has an exact electro-optic feedback counterpart, at 
least as far as the effect on the source cavity is concerned. If $V$ 
depends on the intensity of the driven cavity, then the result can be 
reproduced by feedback based on direct photodetection, while 
quadrature-dependent all-optical feedback can be reproduced by 
homodyne detection. The explanation for this is that the nonlinear 
interaction acts as an effective measurement of the driven cavity 
intensity or 
quadrature, which (because it is heavily damped) is a measurement of 
the intensity or quadrature of the output of the first cavity. Unlike the 
electro-optic case, the measuring device then directly influences the 
source cavity through the same coupling, instead of having all of the 
intermediate equipment. On the other hand, a purely optical feedback 
scheme which is sensitive to both quadratures of the driven cavity 
cannot be replicated by currents. That is because such an interaction is 
not like a simultaneous measurement of both quadratures of the driven 
cavity. Rather, the complex amplitude of the driven cavity interacts 
coherently with another non-Hermitian quantity in the source cavity. 
An analogous electro-optic scheme using heterodyne detection can be 
defined, which has the same semiclassical effect as the complex 
amplitude feedback. However, the two independent measurements of 
the two quadratures make the feedback incoherent and introduces 
extra diffusion terms into the master equation.

The equivalence of electro-optic and all-optical feedback for the 
intensity and quadrature feedback means that the latter is subject to 
the same restrictions as the former as far as the generation of 
nonclassical light is concerned. Specifically, if the nonlinear 
interaction 
Hamiltonian $V$ is linear in the {\em source} cavity amplitude or 
intensity, then the feedback cannot produce a nonclassical state in the 
source cavity. The electro-optic equivalent of this theorem is that 
feedback based on extra-cavity detection cannot produce nonclassical 
light by driving or detuning the cavity. However, there is a property of 
all-optical feedback  not present in electro-optical schemes which 
makes the situation less clear cut: an output beam. This is the 
feedback loop beam reflected from the second cavity. It turns out that 
this beam may be arbitrarily squeezed, even though the source 
cavity remains classical. 

The way to understand this is using the measurement analogy 
explained above. The driven cavity is a QND 
measurement device for say the $x$ quadrature of the output of the 
source cavity. This device also may control the dynamics of the 
source cavity such that the measured fluctuations in the $x$ 
quadrature are suppressed. Thus the $x$ quadrature of the output of 
the 
source cavity can be squeezed; its statistics are unchanged upon 
reflection 
at the second cavity. However, it is not until the output reflects off the 
second cavity (the QND device) that the extra fluctuations in the $y$ 
quadrature are put in, due to the effect of the measurement. This 
would seem to indicate that in the loop, Heisenberg's uncertainty 
relations fail, even though they are satisfied in the output loop (which 
is all that is observable). However, this is not the case. The canonical 
commutation relations, which are actually between different spatial 
points of the field at the same time, are never violated. The 
anticorrelations in the output field which result in squeezing are only 
present between parts of the field which are separated in time by more 
than the time delay in the feedback loop. There is no contradiction.

Thus, it would seem that there is one potential application for 
all-optical feedback: producing squeezed light. However, in order to 
build such a device producing well-squeezed light, the coupling would 
have to use a frequency- (but not polarization-) degenerate 
$\chi^{(2)}$ crystal, as well as a polarization converter.
It would seem  easier to use the traditional squeezer, a 
$\chi^{(2)}$ crystal acting as a degenerate parametric oscillator. 
Intensity-dependent  all-optical feedback is even less practical, 
requiring a low loss $\chi^{(3)}$ nonlinearity to operate. The 
smallness of higher order nonlinearities is sufficient justification as to 
why we have not considered all-optical feedback with a coupling 
dependent on higher order field moments of the driven cavity. In fact, 
such higher order feedback does not produce any new results. At least 
in the regime where the second cavity can be adiabatically eliminated, 
the higher order terms either give a vanishing contribution, or 
reproduce the results of amplitude or intensity feedback. Thus we can 
conclude that all-optical feedback is probably not a practical way of 
controlling quantum noise, although there may be other applications. 
Nevertheless, the predicted results are interesting, and some 
experiments should be feasible with current technology. The 
similarities 
with and differences from electro-optic feedback yield important 
insights about the nature of feedback in general.

\acknowledgements
We wish to acknowledge fruitful discussions with W.J. Munro and 
C.W. Gardiner, and H.J. Carmichael. Part of this work was done 
while GJM was visiting the Department of Physics, University of 
Oregon, with the support of a University of Queensland International 
Collaborative Research Grant.

\includegraphics{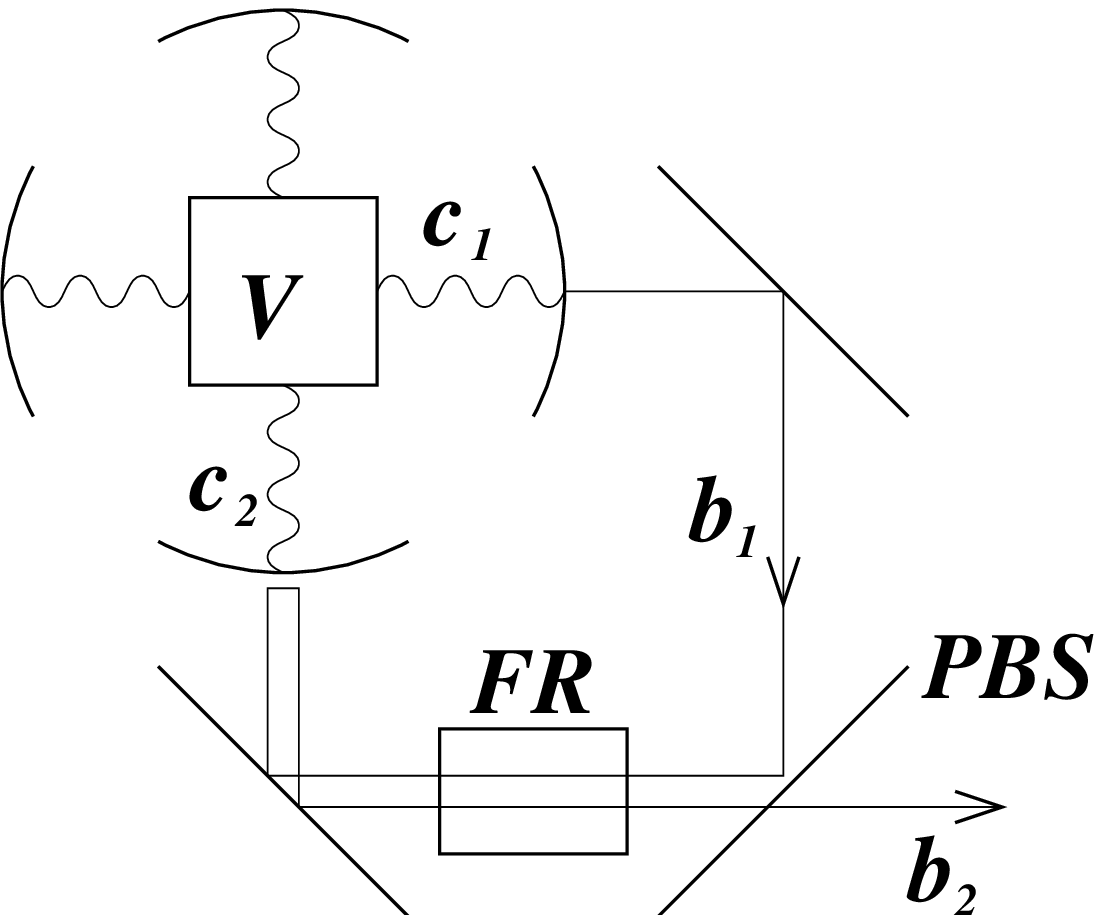}


\begin{thebibliography}{10}

\bibitem{JOSABSS87}
``Squeezed states of the electromagnetic field"
H.J. Kimble and D.F. Walls (eds.),
special issue of J. Opt. Soc. Am. B {\bf 4}, 1453 (1987).

\bibitem{YamImoMac86} 
Y. Yamamoto, N. Imoto and S. Machida, 
Phys. Rev. A {\bf 33}, 3243 (1986).

\bibitem{HauYam86}
H.A. Haus and Y. Yamamoto,
Phys. Rev. A {\bf 34}, 270 (1986)

\bibitem{Sha87}
J.M. Shapiro {\em et al},
J. Opt. Soc. Am. B {\bf 4}, 1604 (1987)

\bibitem{TapRarSat88} %expt - DPO with fb.
P.R. Tapster, J.G. Rarity, and J.S. Satchell
Phys. Rev. A {\bf 37}, 2963 (1988)

\bibitem{WisMil93b}
H.M. Wiseman and G.J. Milburn,
Phys. Rev. Lett. {\bf 70}, 548 (1993)

\bibitem{WisMil93d}
%H.M. Wiseman and G.J. Milburn,
%to be published in Phys. Rev. A %AA5089
%\bibitem{WisMil94a} %svf
H. M. Wiseman and G. J. Milburn,
%``Squeezing via feedback''
Phys. Rev. A {\bf 49}, 1350 (1994).

\bibitem{Wis93b}
%H.M. Wiseman,
%to be published in Phys. Rev. A. %LD5068AR
%\bibitem{Wis94a} %qtocf
H. M. Wiseman,
%``Quantum theory of continuous feedback''
Phys. Rev. A {\bf 49}, 2133 (1994);
Errata {\em ibid.}, {\bf 49} 5159 (1994) and {\em ibid.} {\bf 50}, 
4428 (1994).

\bibitem{Car93b}
H.J. Carmichael,
{\em An Open Systems Approach to Quantum Optics},
(Springer-Verlag, Berlin, 1993)

\bibitem{DalCasMol92}
J. Dalibard, Y. Castin and K. M\o lmer,
Phys. Rev. Lett. {\bf 68}, 580 (1992).

\bibitem{GarParZol92}
C.W. Gardiner, A.S. Parkins, and P. Zoller,
Phys. Rev. A {\bf 46}, 4363 (1992).

\bibitem{WisMil93c}
H.M. Wiseman and G.J. Milburn,
Phys. Rev. A {\bf 47}, 1652 (1993) 

\bibitem{Gar93}
C.W. Gardiner,
Phys. Rev. Lett. {\bf 70}, 2269 (1993)

\bibitem{Car93a}
H.J. Carmichael,
Phys. Rev. Lett. {\bf 70}, 2273 (1993)

\bibitem{GarCol85}
C.W. Gardiner and M.J. Collett,
\newblock  Phys. Rev. A {\bf 31}, 3761 (1985).

\bibitem{Car87}
H.J. Carmichael,
\newblock J. Opt. Soc. Am. B  {\bf 4}, 1588 (1987).

\bibitem{Bar86}
A. Barchielli,
Phys. Rev. A {\bf 34}, 1642 (1986).

\bibitem{WisMil93a}
H.M. Wiseman and G.J. Milburn,
Phys. Rev. A {\bf 47}, 642 (1993)

\bibitem{Gar91}
C.W. Gardiner,
{\em Quantum Noise}
(Springer-Verlag, Berlin, 1991)

\bibitem{Gar85}
C.W. Gardiner,
{\em Handbook of Stochastic Methods 2e}
(Springer-Verlag, Berlin, 1985)

\bibitem{AlsMilWal88}
P. Alsing, G.J. Milburn, and D.F. Walls,
\newblock Phys. Rev. A {\bf 37}, 2970 (1988).

\end{thebibliography}
\end{document}